\documentclass[journal]{IEEEtran}
\usepackage{xcolor,soul,framed}%,caption
\colorlet{shadecolor}{yellow}
\usepackage[pdftex]{graphicx}
\graphicspath{{../pdf/}{../png/}{../jpg/}}
\DeclareGraphicsExtensions{.pdf,.png,.jpg}
\usepackage[cmex10]{amsmath}
\usepackage{array}
\usepackage{booktabs}
\usepackage{mdwmath}
\usepackage{mdwtab}
\usepackage{eqparbox}
\usepackage{url}
\usepackage{flushend}
\usepackage{tikz}
\usetikzlibrary{plotmarks}
\usetikzlibrary{arrows}
\usetikzlibrary{arrows.meta}
\usetikzlibrary{patterns}
\usepackage{pgfplots}
\usepgfplotslibrary{patchplots}
\pgfplotsset{compat=newest}
\pgfplotsset{plot coordinates/math parser=false}
\pgfplotsset{tick label style={font=\footnotesize},
						 label style={font=\footnotesize},
						 legend style={font=\footnotesize},}
\usepackage{grffile}
\newlength\figureheight
\newlength\figurewidth
\begin{document}
\bstctlcite{IEEEexample:BSTcontrol}

\title{CD and PMD Effect on Cyclostationarity-Based Timing Recovery for
Optical Coherent Receivers}

\author{Dawei Wang, Meng Qiao, Kunjian Lian, and Zhaohui Li%
\thanks{This work was supported in part by the National Key Research and
Development Program of China (2019YFB1803502), in part by the National
Natural Science Foundation of China (U2001601), and in part by the
Key-Area Research and Development Program of Guangdong Province
(2018B010114002, 2020B0303040001). (\emph{Corresponding author: Dawei
Wang})}%
\thanks{M. Qiao and K. Lian are with Guangdong Provincial Key Laboratory
of Optoelectronic Information Processing Chips and Systems, Sun Yat-sen
University, 510275, Guangzhou, China (e-mails:
qiaom@mail2.sysu.edu.cn; liankj@mail2.sysu.edu.cn).}%
\thanks{D. Wang and Z. Li are with Guangdong Provincial Key Laboratory of
Optoelectronic Information Processing Chips and Systems, School of
Electronics and Information Technology, Sun Yat-sen University, and also
with Southern Marine Science and Engineering Guangdong Laboratory
(Zhuhai), 519000, Zhuhai, China (e-mails: wangdw9@mail.sysu.edu.cn;
lzhh88@mail.sysu.edu.cn).}}

\markboth{Manuscript Submitted to Journal of Lightwave Technology on \today}{}

\maketitle

\begin{abstract}
Timing recovery is critical for synchronizing the clocks at the
transmitting and receiving ends of a digital coherent communication
system. The core of timing recovery is to determine reliably the current
sampling error of the local digitizer so that the timing circuit may
lock to a stable operation point. Conventional timing phase detectors
need to adapt to the optical fiber channel so that the common effects of
this channel, such as chromatic dispersion (CD) and polarization mode
dispersion (PMD), on the timing phase extraction must be understood.
Here we exploit the cyclostationarity of the optical signal and derive a
model for studying the CD and PMD effect. We prove that the CD-adjusted
cyclic correlation matrix contains full information about timing and
PMD, and the determinant of the matrix is a timing phase detector immune
to both CD and PMD. We also obtain other results such as a completely
PMD-independent CD estimator, etc. Our analysis is supported by both
simulations and experiments over a field implemented optical cable.
\end{abstract}

\begin{IEEEkeywords} timing recovery,
polarization mode dispersion, 
signal processing.
\end{IEEEkeywords}

% % For peer review papers, you can put extra
% information on the cover % page as needed:
% \begin{center}
% \bfseries EDICS Category: 3-BBND
% \end{center}
% \fi
% %
% % For peerreview papers, this IEEEtran command
% inserts a page break and % creates the second title.
% It will be ignored for other modes.
% \IEEEpeerreviewmaketitle

\section{Introduction}
\IEEEPARstart{T}{iming} recovery plays a critical role in the digital
signal processing (DSP) chain of optical coherent receivers for
synchronizing the local clock to the incoming data stream
\cite{savory2010digital,faruk2017digital,kikuchi2015}. The core of
timing recovery is a timing phase detector (TPD), which identifies the
current sampling phase of the receiver analog-to-digital convertor (ADC)
\cite{zhou2014eff}, often equivalent to detecting the phase of a
generated clock tone at the frequency equal to the signal baudrate. The
core can also be a timing error detector (TED), which produces error
signals indicating the deviation of current sampling instance from the
optimal sampling point. While finding a proper TPD or TED is a classic
problem in digital communication systems, and there are numerous methods
for solving it \cite{godard1978,gardner1986,oerder1988}, one should pay
attention to various fiber channel effects on the existing timing
recovery solutions in optical communications \cite{diniz2018clock}. Such
effects include fiber chromatic dispersion (CD), state of polarization
(SOP) rotation, polarization mode dispersion (PMD), and fiber
nonlinearity, etc. Considerable efforts have been made to understand the
CD and PMD effect on the timing recovery
\cite{hauske2010impact,zibar2011exp}. It appears however that there are
still more to discover especially concerning the spectral correlation
involving the two data-carrying polarizations. Both CD and PMD effects
can be accounted for and simple TED algorithms can be derived from the
the so-called cyclic correlation matrix. In particular, we prove that
the determinant of the proper matrix is a valid TPD regardless of the CD
and PMD conditions of optical fiber and hence can be performed before CD
and PMD equalization in the DSP chain. More analysis on the spectral
correlation properties of the common timing recovery methods can also
found in \cite{huang2014perf,wang2021modified}.

The effect of CD on timing recovery is static and hence it is a common
practice to extract timing phase after CD equalization. Because the
clock tone is sensitive to CD, it also becomes a popular way to estimate
the CD based on the magnitude of clock tone \cite{malouin2012natural}.
In contrast, the PMD effect is dynamic and the half symbol first-order
differential group delay (DGD) rotated by 45-degree is well identified
as the worst case for timing recovery. In fact, based on our analysis in
this paper, the worst cases correspond to the slow principal state of
polarization (PSP) of optical fiber lying in the $s_2$-$s_3$ plane of
Stokes space (Section \ref{sec:iiia}). It then follows that one could
rotate the signal polarization away from the worst case scenario prior
to the timing recovery. Previous studies include adaptive polarization
rotation to maximize the time-averaged error signals of Gardner TED
\cite{zibar2011exp}, adaptive phase and/or polarization rotation to
maximize cost functions similar as the clock-tone strength
\cite{sun2011novel,stojanovic2012}, SOP rotation to minimize the
correlation between two polarizations \cite{rozental2017}. Other timing
phase extraction points such as the output of channel equalizer and
carrier recovery are discussed in \cite{sun2015clock}.

It has been understood that the PMD effect is related to the
polarization correlation. However, we haven't seen a complete analysis
on what exactly the relation is and its impact on the timing recovery.
In this paper, we prove that the spectral correlation and the cyclic
correlation functions are powerful tools for analyzing the channel
effects on the timing recovery. We analyze in detail the CD and PMD
effect on the class of timing error detectors based on the second-order
cyclostationarity of data-carrying signals. We have proposed several
algorithms in that class including the a completely PMD-insensitive CD
estimator, a PMD matrix estimator, the DGD and PSP vector estimation,
and the TED for negligible DGD, for DGD smaller that half unit interval,
and for all channel conditions. We corroborate our findings with
numerical simulations of an optical coherent communication system
exploiting the polarization division multiplexed 16-ary quadrature
amplitude modulation (PDM-16QAM) and field tests with a 153 km
subterranean optical cable.

\section{Principle of Timing Recovery}
\begin{figure}[t]
\centering
\includegraphics[width=\columnwidth]{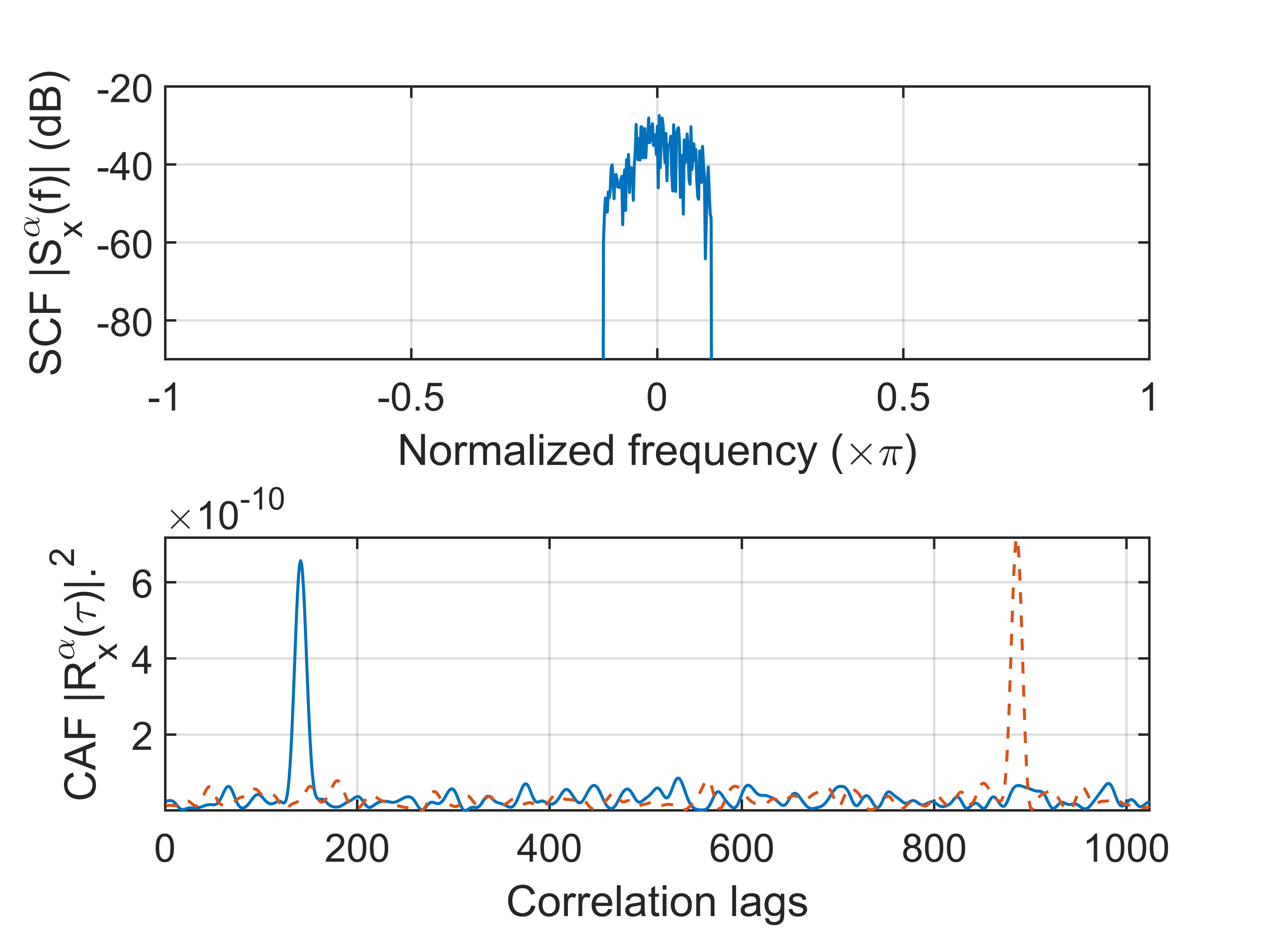}
\caption{Simulation results of SCF and CAF based on a 32 GBaud PDM-16QAM
system with the cyclic frequency equal to the baudrate, 18 dB OSNR, 100
kHz laser linewidth, $\pm$8.5 ns/nm CD, and 512 symbols. The CAF curve
of dashed line is for the CD with negative sign.}
\label{f1}
\end{figure}
Consider the general linear modulation carried by one of two orthogonal
polarizations ($x$ polarization) of light
\begin{equation}
x(t)=\sum_{n=-\infty}^{\infty} a_n g(t-\tau_g-nT_0)
\end{equation}
where $\{a_n\}$ are the complex-valued symbols randomly selected from a
signal constellation, $T_0$ is one symbol duration, $g(t)$ is the
real-valued pulse shaping function, and $\tau_g$ is the group delay
causing the varying timing phase. The signal $y(t)$ at the $y$
polarization is defined similarly but with symbols $\{b_n\}$ that are
uncorrelated with $\{a_n\}$. The cyclic autocorrelation function (CAF)
of $x$ is given by
\begin{align}
\label{eq:caf}
R_x^{\alpha}(\tau) &= \frac{1}{T_0} \int_{-T_0/2}^{T_0/2}
E\left\{ x\left(t+\frac{\tau}{2}\right)x^*\left(t-\frac{\tau}{2}\right)
e^{-j2\pi\alpha t} \right\} \,dt\\
\label{eq:caf2}
&= \frac{1}{T_0^2} \int_{-\infty}^{\infty}
g\left(t-\tau_g+\frac{\tau}{2}\right) g\left(t-\tau_g-\frac{\tau}{2}\right)
e^{-j2\pi\alpha t} \,dt
\end{align}
where $j=\sqrt{-1}$, $E\{\cdot\}$ is the expectation operator and the
cyclic frequency $\alpha$ is an integer multiple of the signal baudrate.
The spectral correlation function (SCF) of $x$ can be found as the
Fourier transform of the CAF
\begin{align}
\label{eq:scf}
S_x^{\alpha}(f)
&= \int_{-\infty}^{\infty} R_x^{\alpha} e^{-j2\pi f\tau}
\,d\tau
\\
\label{eq:scf2}
&= \frac{1}{T_0^2}
G\left(f+\frac{\alpha}{2}\right)
G^*\left(f-\frac{\alpha}{2}\right)
e^{-j2\pi\alpha\tau_g}\,,
\end{align}
which can be viewed as a generalization of ordinary power spectral
density (a special case of SCF with $\alpha =0$). When the pulse shaping
$g(t)$ is an even and real function, its Fourier transform $G(f)$ is
also even and real. Also, when $G(f)$ is bandlimited with a nonzero
excess bandwidth, the SCF is nonzero only for a few cyclic frequencies.
Fig. \ref{f1} shows simulated SCF and CAF curves for a common modulation
format used in coherent optical systems. The fiber CD caused the shift
of CAF peak, which is explained in more detail in Section IV.
Moreover, it is easy to verify the spectral correlation is perfect for
common digital modulations in the sense that the correlation coefficient
$S_x^{\alpha}(f)/\sqrt{S_x(f+\alpha/2)S_x(f-\alpha/2)}$ has modulus one
for $\alpha \neq 0$. The spectral components that are separated by
$\alpha$ are effectively carrying the same information. Note that in
contrast to the real-valued power spectral density, the SCF is in
general complex-valued due to the presence of phase term involving
$\tau_g$ for $\alpha \neq 0$. This is the reason that enables us to
retrieve the timing phase term by averaging the SCF with $\alpha
=1/T_0$, the baudrate, over a certain range of frequency. That is $\int
S_x^{\alpha} \,df = A\cdot e^{-j2\pi\alpha\tau_g}$ according to
(\ref{eq:scf2}), where $A$ is a real-valued coefficient. The integration
range is usually small around the zero frequency such that we compute
effectively the correlation between upper and lower bands of the signal
around $f=\pm 1/(2T_0)$. It is obvious that the real and the imaginary
part follows
\begin{align}
\operatorname{Re}{\int S_x^{\alpha} \,df} &= A\cdot \cos(2\pi\alpha\tau_g)\\
\operatorname{Im}{\int S_x^{\alpha} \,df} &= -A\cdot \sin(2\pi\alpha\tau_g)
\end{align}
where the imaginary part forms the well-known timing error indicator (or
detector) of Godard's method \cite{godard1978}. By using (\ref{eq:scf}),
we obtain easily the timing error detector
\begin{align}
\label{eq:ted}
e_T &= -\operatorname{Im}{\int S_x^{\alpha} \,df} 
= -\operatorname{Im}R_x^{\alpha}(0)\\
\label{eq:ted2}
&= -\operatorname{Im}\left\{ 
\frac{1}{T}\int_{-T/2}^{T/2} x(t)x^*(t)e^{-j2\pi\alpha t} \,dt
\right\}
\,,
\end{align}
where we have replaced the expectation with time average over a large
$T$. The last line is recognized as the square timing error detector
\cite{oerder1988} which computes the complex coefficient of the spectral
line of $|x|^2$ at frequency $\alpha =1/T_0$. It can be shown
\cite{gardner1986role} that in order to maximize the SNR of the spectral
line, $x(t)$ needs to be the matched filtered signal at the receiver.
The TED is usually embedded in a feedback loop which adjusts the ADC
sampling phase constantly to lock on the optimal timing instances.

% quadratic time-invariant transformation

\section{PMD Effect}
The formulation (\ref{eq:caf}--\ref{eq:scf2}) works equally well for the
$y$ polarization. Also, the cyclic cross-correlation function (CCF) and
the SCF between the two orthogonal polarizations can be similarly
developed. They can be written as matrices of form
\begin{align}
\label{eq:csmtx}
C(\tau) = \begin{bmatrix}
R_{xx}^{\alpha} & R_{xy}^{\alpha}\\
R_{yx}^{\alpha} & R_{yy}^{\alpha}
\end{bmatrix}
\qquad
S(f) = \begin{bmatrix}
S_{xx}^{\alpha} & S_{xy}^{\alpha}\\
S_{yx}^{\alpha} & S_{yy}^{\alpha}
\end{bmatrix}
\end{align}
where $R_{yx}^{\alpha}$ and $R_{xy}^{\alpha}$ are given respectively by
replacing in (\ref{eq:caf}) the first and the second $x(t)$ by $y(t)$.
In ideal cases (i.e., no CD, no PMD, and stationary noise), the $S$
matrix is (up to a common phase term) an identity matrix, $S=
e^{-j2\pi\alpha\tau_g} I$, at a given frequency because the data
carried by $x$ and $y$ polarization are uncorrelated. Hence, according
to (\ref{eq:ted}), the CAF matrix
\begin{align}
\label{eq:c0}
C(0) = \int S(f)\,df = Ae^{-j2\pi\alpha\tau_g} I
\,,
\end{align}
where $A$ is a real-valued coefficient.
\begin{figure}[t]
\centering
\includegraphics{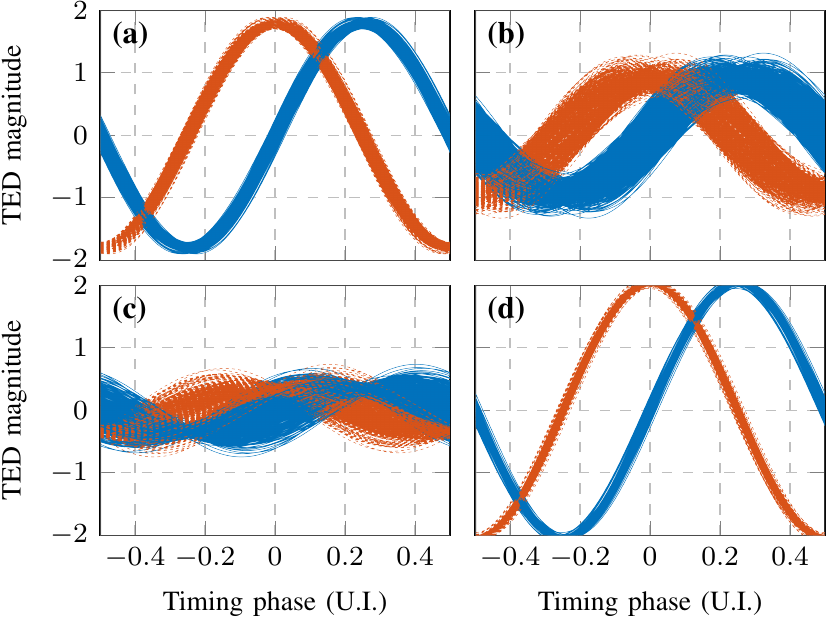}
\caption{Characteristic curves of timing error detector derived from (a)
$\operatorname{trace}(\bar{P'})$ [equ. (\ref{eq:ted_smalldgd})] with $5$
ps DGD, (b) $P'_{xx}$ with $5$ ps DGD, (c)
$\operatorname{trace}(\bar{P'})$ [equ. (\ref{eq:ted_smalldgd})] with
$14$ ps DGD, and (d) $\operatorname{trace}(\bar{P'U^H})$ [equ.
(\ref{eq:ted_largedgd})] with $14$ ps DGD. Solid lines: Imaginary part,
$-\operatorname{Im}\{\cdot\}$. Broken lines: Real part,
$\operatorname{Re}\{\cdot\}$. Simulation parameters: $32$ GBaud
PM-16QAM, $20$ dB OSNR, $100$ kHz laser linewidth with no carrier
frequency offset, FFT size $1024$ samples ($2$ samples per symbol),
random polarization rotation.}
\label{fig:ted_trace1}
\end{figure}
When considering only the first-order PMD of optical fiber with no CD,
we describe the change of output polarization state $E$ given the same
input state over angular frequency $\omega$ by the first-order
differential equation
\begin{align}
% \label{}
\frac{dE}{d\omega} = -j(\tau_g + H)E
\,,
\end{align}
where the matrix $H$ is Hermitian when the polarization dependent loss
(PDL) is neglected. It represents the rotation of output Stokes vector
about the principal state of polarization (PSP) when frequency changes.
The rotation rate is precisely the differential group delay (DGD). The
input polarization states are the same at the two frequencies
$f\pm\alpha/2$ due to the complete spectral correlation. Therefore, when
neglecting PDL, a constant unitary matrix $U$ (PMD matrix) relates the
spectral components as follows
\begin{align}
\label{eq:pmd}
\underbrace{\begin{bmatrix} X(f+\alpha/2)\\ Y(f+\alpha/2)
\end{bmatrix}}_{E'_1} = e^{-j\phi_0} U
\underbrace{\begin{bmatrix} X(f-\alpha/2)\\ Y(f-\alpha/2)
\end{bmatrix}}_{E'_2}\,,
\end{align}
where $\phi_0=2\pi\alpha\tau_g$ is the common phase induced by the group
delay, $E'_{1,2}$ are Jones vectors, $X$ and $Y$ are respectively the
Fourier transform of time truncated $x(t)$ and $y(t)$ at the receiver.
The PMD matrix can be written in the form
\begin{align}
\label{eq:uni}
U = \begin{bmatrix}
a & -b^*\\
b & a^*
\end{bmatrix} \quad \text{with } aa^*+bb^*=1\,,
\end{align}
and admits the expansion
\begin{align}
\label{eq:uex}
U = \cos(\pi\alpha\tau_\text{DGD})I 
-j\sin(\pi\alpha\tau_\text{DGD})[\vec{p}\cdot\vec{\sigma}]
\,,
\end{align}
where $\vec{p}=[p_1,p_2,p_3]$ is the slow PSP Stokes vector associated
with the eigenvalue $\rho = e^{-j\pi\alpha\tau_\text{DGD}}$ of $U$,
$\vec{\sigma}=[\sigma_1,\sigma_2,\sigma_3]$ are the three Pauli matrices
\begin{align}
\label{eq:pauli}
\sigma_1 = \begin{bmatrix}1&0\\0&-1\end{bmatrix},
\sigma_2 = \begin{bmatrix}0&1\\1&0 \end{bmatrix},
\sigma_3 = \begin{bmatrix}0&-j\\j&0\end{bmatrix}\,,
\end{align}
and $\vec{p}\cdot\vec{\sigma} = p_1\sigma_1 + p_2\sigma_2 +
p_3\sigma_3$,
\begin{figure}[t]
\centering
\includegraphics{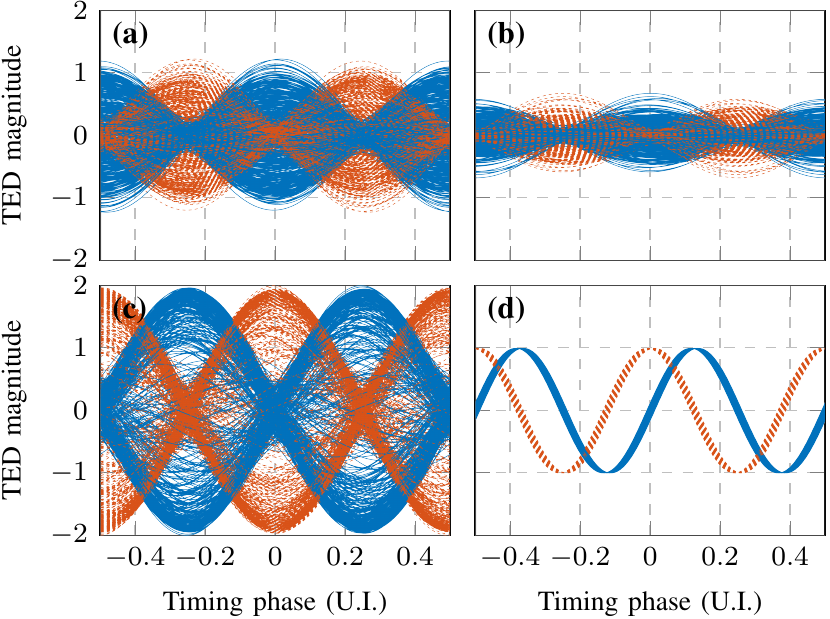}
\caption{Characteristic curves of timing error detector derived from (a)
$P'_{xx}$, (b) $\operatorname{trace}(\bar{P'})$ [equ.
(\ref{eq:ted_smalldgd})], (c) $\operatorname{trace}(\bar{P'U^H})$ [equ.
(\ref{eq:ted_largedgd})], and (d) $\operatorname{det}(\bar{P'})$ [equ.
(\ref{eq:cpd})] when the DGD is $15.63$ ps. Solid lines: Imaginary part,
$-\operatorname{Im}\{\cdot\}$. Broken lines: Real part,
$\operatorname{Re}\{\cdot\}$. Other simulation conditions are the same
as those for Fig. \ref{fig:ted_trace1}.}
\label{fig:ted_det}
\end{figure}
\subsection{Cyclic Periodogram}
\label{sec:iiia}
Let $E_{1,2}$ be the two output Jones vectors at frequencies
$f\pm\alpha/2$ when this is no PMD and polarization rotation. Namely,
they should coincide with their input states except for a common phase
change. Using the relations $E_1 = e^{-j\phi_0}E_2$, $E'_2=VE_2$ and
(\ref{eq:pmd}), we obtain the cyclic periodogram
\begin{align}
\label{eq:cp}
P'=E'_1(E'_2)^H=UV(E_1(E_2)^H)V^H=UVPV^H\,,
\end{align}
where $V$ is a unitary matrix describing the global polarization
rotation of the fiber. The matrix $P$ is the cyclic periodogram without
polarization rotation and PMD, which is in analogy to the ordinary
periodogram, a practical estimate of the SCF matrix $S$. By taking
the expectation of both sides of (\ref{eq:cp}), or by averaging both
sides of (\ref{eq:cp}) over all frequencies, we see that
\begin{equation}
\label{eq:pmde}
\hat{C}(0) = \bar{P'} = UV\bar{P}V^H \approx e^{-j\phi_0}U
\end{equation}
because $\bar{P}\approx e^{-j\phi_0}I$ and the first equality is due to
the Fourier relation in (\ref{eq:c0}). We have chosen $A=1$ assuming
proper normalization and $\hat{C}$ is the estimate of $C$ in
(\ref{eq:csmtx}). Namely, the frequency averaged cyclic periodogram
matrix $\bar{P'}$ or the estimated CAF matrix $\hat{C}$ at $\tau=0$ is
in fact an estimate of the PMD matrix $U$, with a common phase related
to the group delay. Matrices $\bar{P'}$ and $\bar{P}$ also have four
elements
\begin{align}
\label{eq:cpmtx}
\bar{P'} = \begin{bmatrix}
P'_{xx} & P'_{xy}\\
P'_{yx} & P'_{yy}
\end{bmatrix}
\qquad
\bar{P} = \begin{bmatrix}
P_{xx} & P_{xy}\\
P_{yx} & P_{yy}
\end{bmatrix}\,.
\end{align}
% Note that the PMD matrix estimate given by (\ref{eq:pmde}) is
% insensitive to the common polarization rotation but is affected by the
% timing error.
For $U$ given by (\ref{eq:uex}), it is easily seen that $\bar{P'} = \pm
je^{-j\phi_0}[\vec{p}\cdot\vec\sigma]$ whenever $\tau_\text{DGD} =
0.5\alpha^{-1} \operatorname{mod} (\alpha^{-1})$, commonly known as the
half symbol DGD, and furthermore $P'_{xx} = P'_{yy} = 0$ whenever $p_1 =
0$, i.e., the Stokes vector of fiber slow PSP has zero $s_1$ component.
This defines the worst cases for exploiting any linear combination of
$P'_{xx}$ and $P'_{yy}$ as TED.

% We will denote $\bar{P'}$ and $\hat{C}(0)$ by $\widehat{U_T}$.

\subsection{Case of Small DGD}
When the DGD is negligible such as in short reach fiber communications,
the PMD matrix $U$ is approaching an identity matrix such that
\begin{align}
\label{eq:ted_smalldgd}
\operatorname{trace}(\bar{P'}) =
\operatorname{trace}(V\bar{P}V^H) \approx 2 e^{-j\phi_0}
\,,
\end{align}
the imaginary part of which,
$-\operatorname{Im}\{\operatorname{trace}(\bar{P'})\}$, is a valid
timing error detector completely independent of the polarization
rotation due to the $\operatorname{trace}$ operation. Note that since
$\bar{P}$ is basically a noisy identity matrix, the first element of
$\bar{P'}$, the $P'_{xx}$ in (\ref{eq:cpmtx}), should also be a timing
detector. However, it can be shown that the SNR of $P'_{xx}$ is lower
than that of $\operatorname{trace}(\bar{P'}) = P_{xx} +P_{yy}$, leading
to a worse jitter performance. This is because $P'_{xx}$ in fact
involves all four elements of $\bar{P}$ and does not cancel completely
the effect of $V$.
\begin{figure}[t]
\centering
\includegraphics[width=0.9\columnwidth]{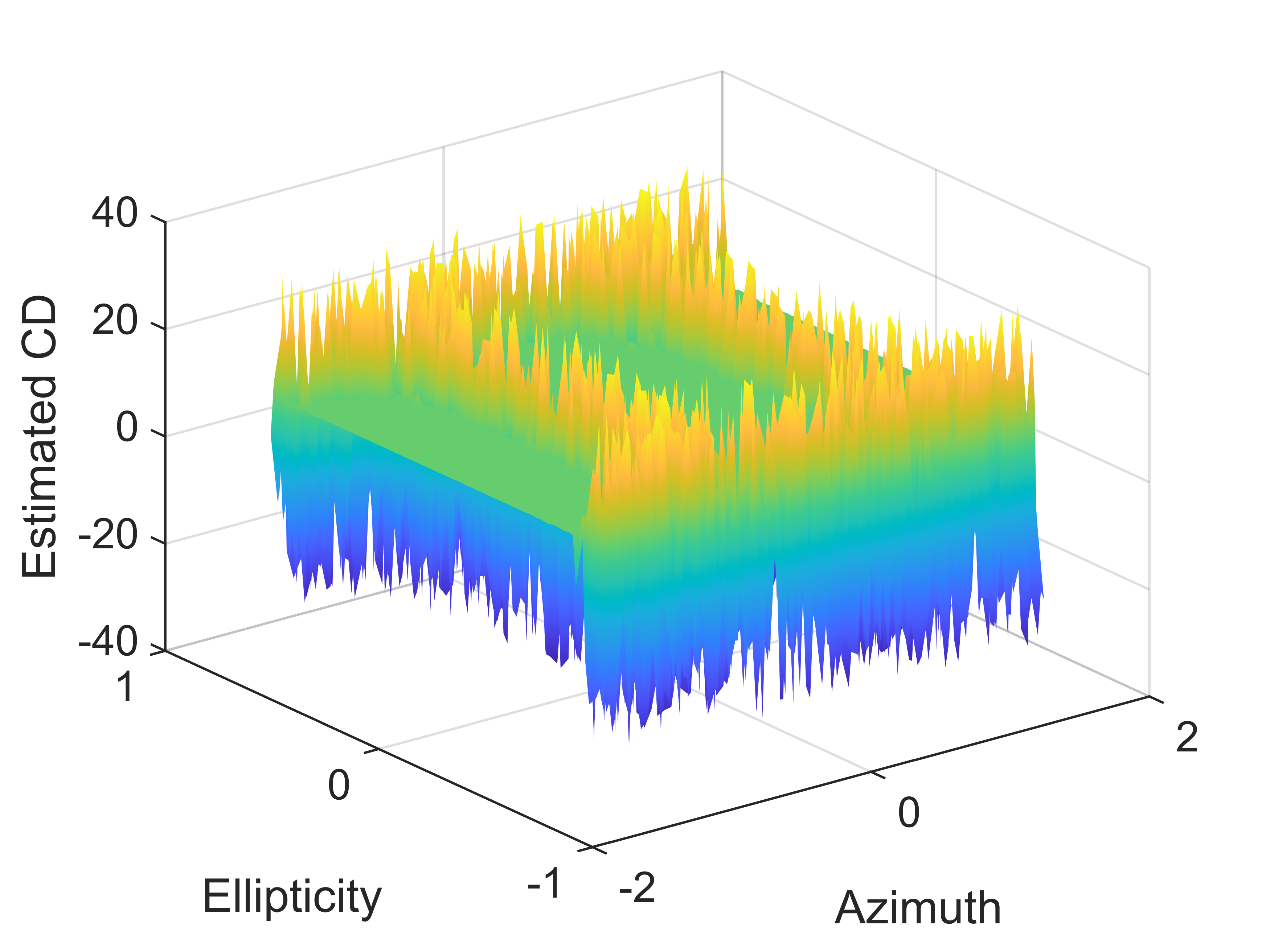}
\includegraphics[width=0.9\columnwidth]{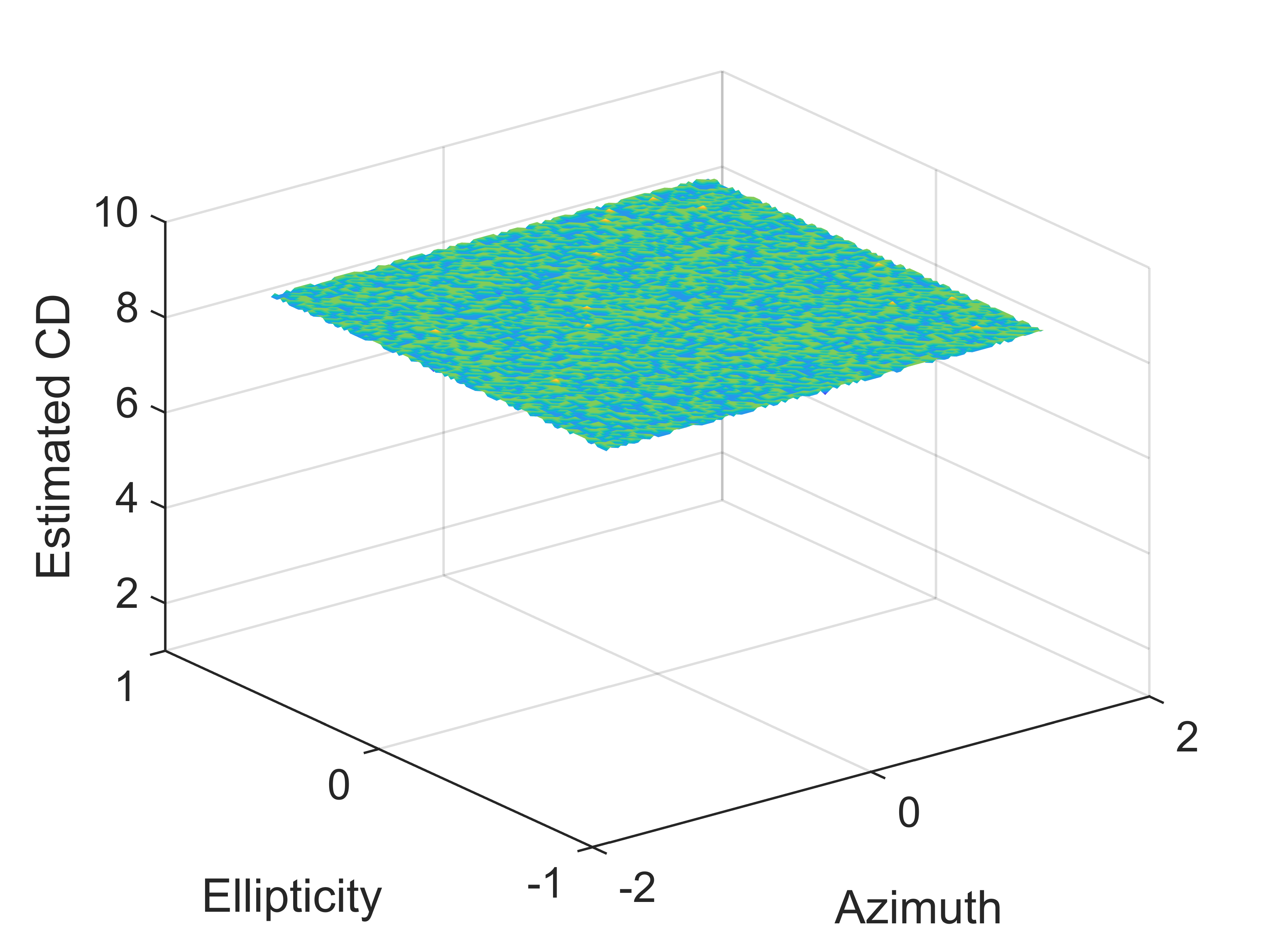}
\caption{Simulation results of CD estimation (ns/nm) based on the
estimator given by (a) equ. (\ref{eq:cdxx}) and (b) equ.
(\ref{eq:cdxy}), respectively. Simulation parameters are the same as
those for Fig. \ref{fig:ted_det} with fiber CD of $8.5$ ns/nm.}
\label{fig:cd}
\end{figure}
\subsection{Case of Large DGD}
When the DGD is large, we first note that
$\operatorname{trace}(\bar{P'})$ has a clear dependence on the PMD [cf.
(\ref{eq:pmde})]. But the quantity
\begin{equation}
\label{eq:ted_largedgd}
\operatorname{trace}(\bar{P'}U^H) =
\operatorname{trace}(UV\bar{P}V^HU^H) \approx 2 e^{-j\phi_0}
\end{equation}
is an exact timing phase detector independent of PMD, provided we know
$U$. However, there is no obvious way to estimate the true $U$ without
being affected by the timing phase [cf. (\ref{eq:pmde})]. On the other
hand, it is remarkable that despite the potentially time-varying phase
term $e^{-j\phi_0}$, both DGD and PSP can be estimated from the matrix
$\bar{P'}$. Since $U$ has two eigenvalues $\rho = e^{\pm
j\pi\alpha\tau_\text{DGD}}$, the DGD can be estimated from the two
eigenvalues of $\bar{P'}$, the estimate of $U$, via
\begin{align}
\label{eq:dgde}
\widehat{\tau_\text{DGD}} = |\arg\{\rho_1\rho_2^*\}|/(2\pi\alpha)
\end{align}
with a wrapped range from zero to half symbol duration due to the 
multi-valued argument function $\arg(\cdot)$ and the PSP
vector can be estimated via
\begin{align}
\label{eq:pspe}
\widehat{p_i} = k\cdot\operatorname{Im}\left[
\operatorname{trace}\left(
\sigma_i\bar{P'}
\right)\cdot
\operatorname{trace}
(\bar{P'})^H
\right]
\end{align}
where $i=1,2,3$ and $k$ is a constant such that the estimated vector is
normalized. It is easy to verify that $\widehat{p_i}$ has a dependence
$\sin(2\pi\alpha\tau_\text{DGD})$ on the true DGD $\tau_\text{DGD}$.
We can combine (\ref{eq:dgde}) and (\ref{eq:pspe}) to form an estimate
of $U$ that is independent of timing phase, i.e.,
\begin{align}
\hat{U} = \cos(\pi\alpha\widehat{\tau_\text{DGD}})I 
-j\sin(\pi\alpha\widehat{\tau_\text{DGD}})[\vec{p}\cdot\vec{\sigma}]
\,.
\end{align}
Note that the sign of $\hat{U}$ depends on both $\tau_\text{DGD}$ and
$\widehat{\tau_\text{DGD}}$ and cannot be determined based on
$\widehat{\tau_\text{DGD}}$ alone. This sign ambiguity prohibits the use
of $\operatorname{trace}(\bar{P'}\hat{U}^H)$ as a timing phase detector
for an arbitrary DGD. It does work when the true DGD is strictly less
than half of the symbol duration.

\subsection{All DGD Cases}
On the other hand, we see that the determinant of $P'$
\begin{align}
\label{eq:cpd}
\operatorname{det}(\bar{P'}) = 
\operatorname{det}(UV\bar{P}V^H) = e^{-2j\phi_0}
\end{align}
is PMD independent. Its imaginary part,
$-\operatorname{Im}\{\operatorname{det}(\bar{P'})\}$, is a timing error
detector with characteristic curves shown in Fig. \ref{fig:ted_det}(d).
It has two positive zero-crossings and hence two locking points.
Depending on the initial state, the timing recovery loop may lock to one
of the two points. For instance, when the initial timing phase is
$>0.25$ or $<-0.25$ unit interval (i.e., one symbol), it locks to the
symbol edges. This property combined with the varying polarization
mixing due to PMD would not be a prominent problem if fractional-spaced
channel equalizer is used after the timing unit. Otherwise, it might
affect the DSP performance when the symbol-spaced equalizer is used
afterwards. However, this is a common problem for all TEDs considering
the fact that the concept of optimal sampling phase is complicated by
the combined effect of locking point and PMD. Note that the TED based on
(\ref{eq:cpd}) is completely transparent to the first-order PMD and
polarization rotation. It is indeed slightly more complex than the
simple trace operation given by (\ref{eq:ted_smalldgd}). In light
of (\ref{eq:ted2}) and (\ref{eq:cpd}), we could also extract directly
a clock tone, independent of PMD, at the baudrate from the spectrum
\begin{align}
% \label{}
F = F_{xx}F_{yy} - F_{xy}F_{yx}
\,,
\end{align}
where $F_{xx}$ is the fast Fourier transform (FFT) of $xx^*$, $F_{xy}$
is the FFT of $xy^*$, etc., and $x$, $y$ are signals from two
polarizations after CD compensation.

\subsection{Adaptive Method}
First introduced in \cite{hauske2010impact} and furthered studied in
\cite{sun2011novel}, the linear combination of the elements of matrix
$\bar{P'}$ seems a feasible way to deal with the PMD effect. Consider
the quantity
\begin{align}
% \label{}
q = P'_{xx}+e^{j\phi_{xy}}P'_{xy}
+e^{j\phi_{yx}}P'_{yx}+e^{j\phi_{yy}}P'_{yy}
\,,
\end{align}
the imaginary part of which, $-\operatorname{Im}\{q\}$, can be used as
the timing error detector if the phase vector $[\phi_1,\phi_2,\phi_3]$
is adapted jointly with the timing loop such that the above quantity
converges to a purely real-valued number when the joint loop is locked.
The update at time instance $k$ is written as
\begin{align}
% \label{}
\phi^k_l = \phi^{k-1}_l 
-\mu\operatorname{Im}(e^{j\phi^{k-1}_l}P'_{l}),
\quad l=xy,yx,yy
\end{align}
It effectively treats the four entries of the SCF matrix (\ref{eq:cp})
as independent timing phase detectors. Note that unlike previous timing
error detectors, the timing loop does not lock to the true phase
$\phi_0$ via the adaptive method but to the phase of $P'_{xx}$, which is
$\phi_0$ plus the phase of the first element of $U$. This adds noise to
the timing loop considering $U$ is an erratic effect in practice.

\begin{figure}[t]
\centering
\includegraphics{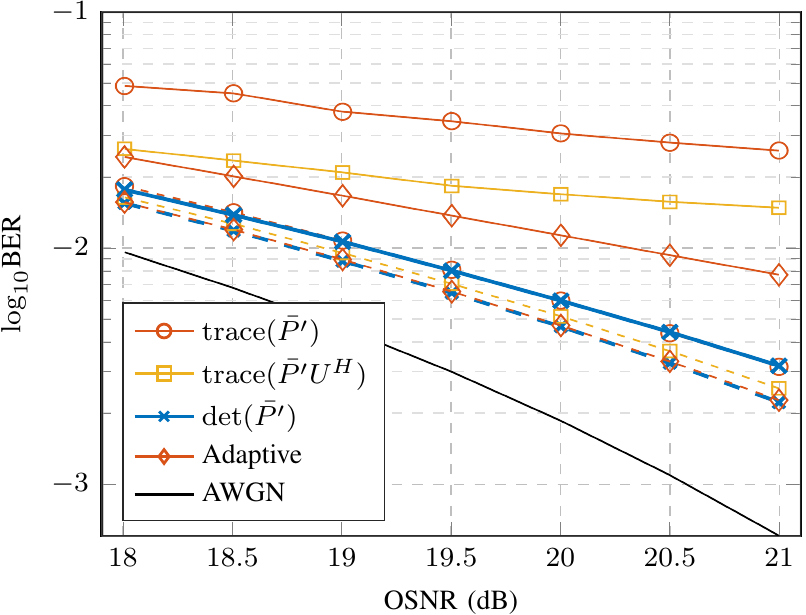}
\caption{Simulation results of a $32$ GBaud PDM-16QAM system bit error
rate (BER) versus optical signal-to-noise ratio (OSNR) when various
timing recovery algorithms are implemented. The solid lines are for
the cases when symbol-spaced LMS equalizer is used. Whereas the dashed
lines are when fractional-spaced equalizer is used.}
\label{fig:sim_ber}
\end{figure}

\section{CD Effect}
It is well-known that the chromatic dispersion of optical fiber incurs a
quadratic phase change over the entire signal spectrum. The frequency
response of CD is given by
\begin{align}
% \label{}
H = \exp(jKf^2)
\end{align}
with $K= \pi\lambda^2DL/c$, where $\lambda$ is the wavelength, $D$
dispersion coefficient, $L$ fiber length, and $c$ the speed of light.
According to (\ref{eq:scf2}), the CD effect results in a linear phase
change in the SCF with the change rate over frequency equal to
$2K\alpha$. The linear phase suppresses the timing error upon
integration in (\ref{eq:ted}). The solution to restore the timing
sensitivity is to estimate the CD value from the slope of phase
\cite{ionescu2015cyclo} and then remove the linear phase from SCF during
integration \cite{sun2011novel}. On the other hand, the linear phase in
SCF is equivalent to a constant time shift in CAF due to their Fourier
relation. The CAF is a pulse-like function and hence the time shift can
be identified by peak searching of $|R_x^{\alpha}|^2$. Namely,
\begin{equation}
\label{eq:cdxx}
\tau_\text{CD} = \arg\max |R_{xx}^\alpha(\tau)|^2
\,.
\end{equation}
Note that the CAF can be computed efficiently by taking the inverse
Fourier transform of SCF. The CAF peak location can also distinguish the
sign of CD, provided the time shift due to CD is no larger than half of
the CAF duration. Moreover, it is easy to verify that the length of CAF
determines the maximum range of CD estimate not the resolution. A higher
resolution of CD estimation can only be obtained by a higher sampling
rate of the signal. After obtaining the delay of CAF peak
$\tau_\text{CD}$, the timing error can be detected simply by using
$\operatorname{Im}R_x^\alpha(\tau_\text{CD})$ \cite{wang2021modified}.
\begin{figure}[t]
\centering
\includegraphics{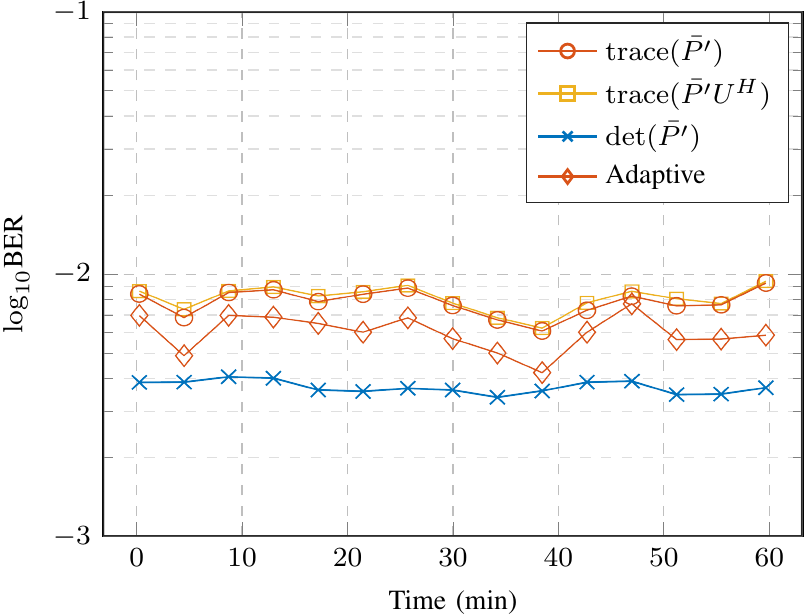}
\caption{Filed-test results of a $1$-hour BER on a $32$ GBaud PDM-16QAM
coherent system over the $153$ km optical cable. A $2000$ rad/s
polarization scrambling at the fiber input is utilized. The DSP setup is
the same as that used for simulations with one sample per symbol, except
that the LMS channel equalizer has $11$ taps for symbol-spaced
configurations. The block length for performing various TEDs is 512
symbols. Each BER result is obtained based on 4 million bits.}
\label{fig:exp_ber}
\end{figure}
Assuming little dependence on polarization, the CD effect is modeled
as a common term in both CAF and SCF. In particular, it causes a time
shift in the CAF matrix and a linear phase change in the SCF matrix
in (\ref{eq:csmtx}). The development in the last section is similar
in the presence of CD,
except that the matrix relation in (\ref{eq:c0}) should be modified as
\begin{align}
\label{eq:ctau}
C(\tau_\text{CD}) = \int S(f) e^{j2\pi f\tau_\text{CD}} \,df 
= Ae^{-j2\pi\alpha\tau_g} I
\end{align}
and in the presence of first-order PMD, $\hat{C}(\tau_\text{CD}) \approx
e^{-j\phi_0}U$ is an PMD matrix estimate insensitive to both CD and
polarization rotation. Note that the form of $U$ given in (\ref{eq:uni})
clearly suggests that the search of CAF peak is affected by the PMD.
Namely, using $|R_{xx}^\alpha|^2$ or $|R_{yy}^\alpha|^2$ or any linear
combination of them will loss sensitivity at $\tau_\text{CD}$ when $a$,
the first element of $U$, is approaching zero. However, the PMD effect
is irrelevant if we use $|R_{xx}^\alpha|^2 + |R_{yx}^\alpha|^2$, because
$|a|^2+|b|^2 = 1$. Namely,
\begin{equation}
\label{eq:cdxy}
\tau_\text{CD} = \arg\max
\big\{|R_{xx}^\alpha(\tau)|^2 + |R_{yx}^\alpha(\tau)|^2\big\}
\,.
\end{equation}
Therefore, we find a way to estimate the CD value completely independent
of polarization rotation, PMD and timing phase. After that, the matrix
$\hat{C}(\tau_\text{CD})$ can be worked with to detect timing error
according to the methods introduced in the last section.

Figure \ref{fig:cd} shows the CD estimation results when (\ref{eq:cdxx})
and (\ref{eq:cdxy}) are used respectively. A half-symbol DGD and all
possible PSP states are considered in the simulation. The results
confirm that (\ref{eq:cdxy}) is insensitive to the PMD. The previous
results suggest that the timing recovery could be performed without
actual CD and/or PMD compensation, which could be useful in certain
practical use cases.

\section{Higher-Order Cyclostationarity}
The studies presented in previous sections are based on second-order
cyclostationarity of common signals, which is however not suitable for
the analysis of severely filtered signal with no excess bandwidth.
Fortunately, the statistics of higher-order cyclostationarity can be
exploited for proper timing phase detection in such cases (to be
published). Although not well recognized by the research community, most
of the proposed timing phase detectors for such signals, such as those
listed in Table 1, are based on the fourth-order cyclostationarity.
However, the CD and PMD effect on this class of timing phase detector
remains unknown. It is not a straightforward task of extending the
analysis in this paper to higher order statistics. It is however clear
that efforts are required to study the spectral properties of
higher-order correlation, which certainly is an interesting direction
for future research.

\begin{table*}[t]
\caption{\textbf{Common fourth-order timing recovery algorithms}}
\centering
\begin{tabular}{cc}
\toprule
Ref. & Algorithm \\
\midrule
\cite{yan2013digital} &
${\sum}_{k=1}^{m}(x_{2k}x_{2k}^*+y_{2k}y_{2k}^*)[(x_{2k+1}x_{2k+1}^*+y_{2k+1}y_{2k+1}^*)-(x_{2k-1}x_{2k-1}^*+y_{2k-1}y_{2k-1}^*)]$\\
\midrule
\cite{stojanovic2014} &
${\sum}_{k=1}^{m}x_{2k}x_{2k+1}^*(x_{2k+1}x_{2k+2}^*-x_{2k-1}x_{2k}^*)+y_{2k}y_{2k+1}^*(y_{2k+1}y_{2k+2}^*-y_{2k-1}y_{2k}^*)$\\
\midrule
\cite{moeneclaey1990} &
${\sum}_{k=1}^{m} x_{k}x^*_{k+1}[x_{k}x^*_{k}-x_{k+1}x^*_{k+1})]$\\	
\midrule
\cite{stojanovic2011}&
$E\left\{\text{Re}\left[(x_{-1}^{*}+x_{0}^{*})(x_{0}+x_{1})((x_{-2}+x_{-1})(x_{-1}^{*}+x_{0}^{*})-(x_{0}+x_{1})(x_{1}^{*}+x_{2}^{*}))\right]\right\}$\\
\bottomrule
\end{tabular}
\end{table*}
\begin{figure}[t]
\centering
\includegraphics[width=0.9\columnwidth]{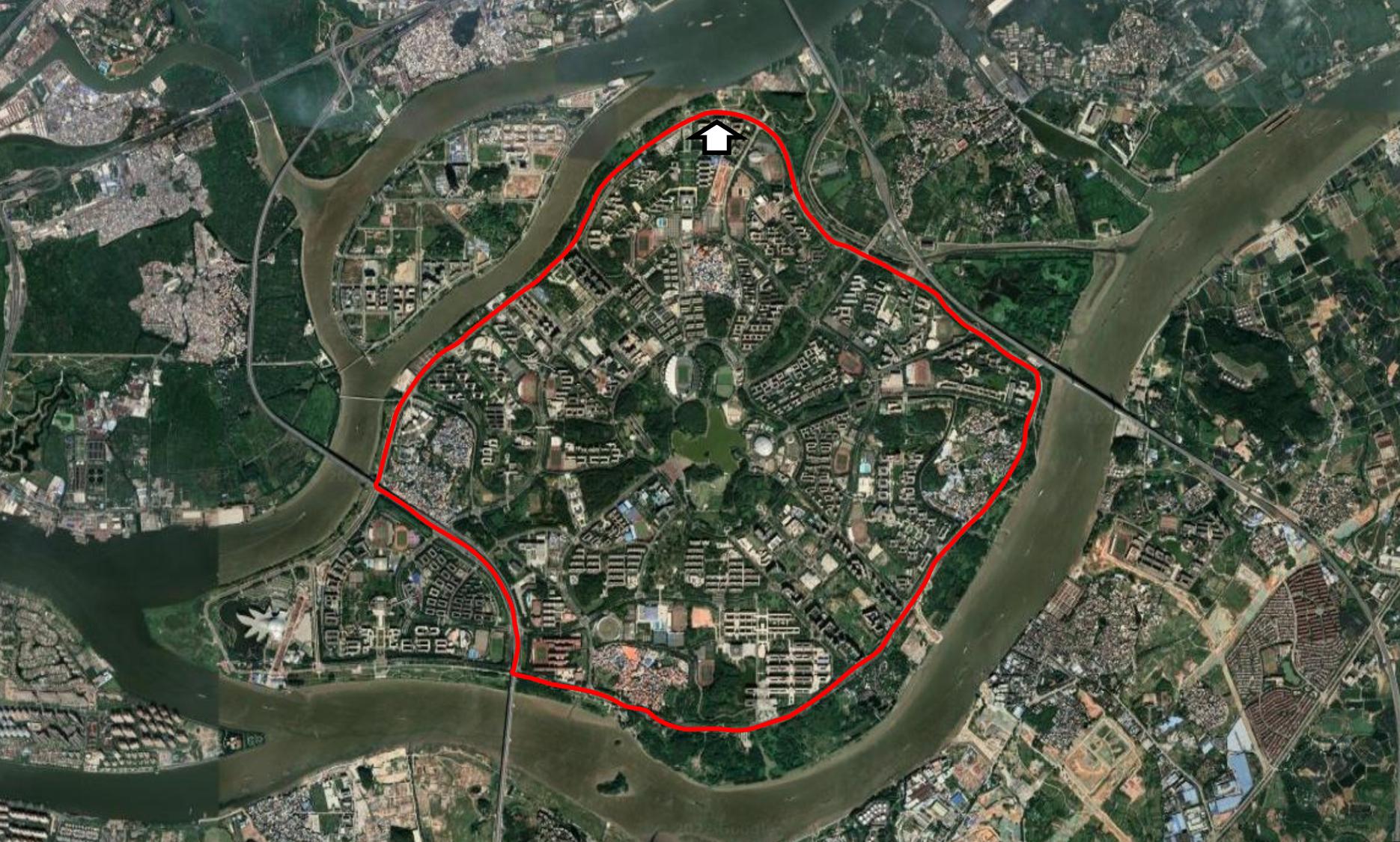}
\caption{The subterranean optical cable containing several single mode
fibers (SMF) used for the field tests. The arrow indicates the location
of our lab and also the two ends of the cable. The optical signal
circulates the cable 9 times, one SMF at a time, and forms the 153 km
optical link for the test.}
\label{fig:field}
\end{figure}
\section{Numerical Simulation}
We perform numerical simulations of a 32 GBaud PDM-16QAM system with 100
km standard single mode fiber (SSMF) transmission. The laser linewidth
is 100 kHz. The laser frequency offset is considered negligible or has
been compensated prior to the timing recovery. The fiber SOP rotation
speed is 50,000 rad/s. The jitter effect of ADC is simulated as a
sinusoidal model with a 30 kHz frequency and $0.6T_0$ peak deviation of
sampling instance. The fiber DGD also varies from 6 ps to 16 ps in a
sinusoidal manner with a frequency of 260 kHz. It provides a critical
situation for testing the performance of various TED algorithms. The DSP
chain includes in sequence CD estimation, CD compensation, timing
recovery, least mean squares (LMS) dual-polarization channel
equalization with embedded phase lock loop (PLL) for carrier phase
recovery. When the LMS equalizer works in the symbol spaced mode, the
performance of timing recovery becomes prominent. The simulation results
are shown in Fig. \ref{fig:sim_ber}. Note that the LMS equalizer has 7
and 13 taps respectively for the symbol-spaced and fractional-spaced
mode for fair comparison. The TED
$-\operatorname{Im}\{\operatorname{det}(\bar{P'})\}$ gives the best
performance for all cases but the differences are less significant when
fractional-spaced equalizer is utilized \cite{kikuchi2011clock}.
\begin{figure}[t]
\centering
\includegraphics[width=\columnwidth]{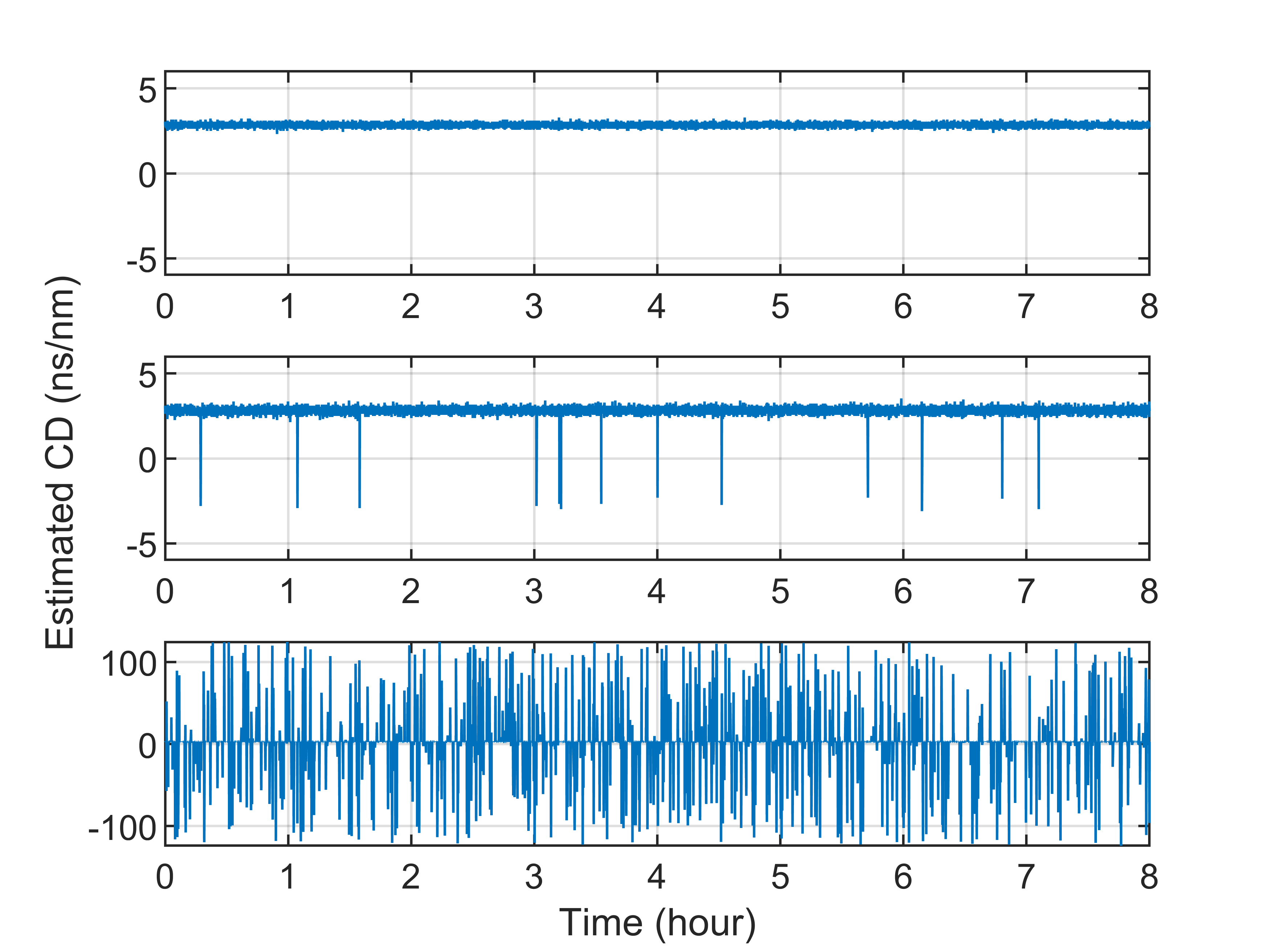}
\caption{Filed-test results of an $8$-hour CD estimation on a $32$ GBaud
PDM-16QAM coherent system over the $153$ km optical cable. Subfigure
from top to bottom: (a) estimator (\ref{eq:cdxy}) with $2000$ rad/s
polarization scrambling at the fiber input, (b) estimator
(\ref{eq:cdxx}) without polarization scrambling, and (c) estimator
(\ref{eq:cdxx}) with $2000$ rad/s polarization scrambling at the fiber
input. Each measurement is performed by evaluating (\ref{eq:cdxx}) and
(\ref{eq:cdxy}) based on a data block containing 2048 symbols.}
\label{fig:exp_cd}
\end{figure}

\section{Field Test}
We exploit the field optical fiber cable as illustrated in Fig.
\ref{fig:field}. It has several standard single mode fibers (SMF) which
can be interconnected at the cable ends so that the signals circulate
inside the fiber loop. The total length of the fiber loop is measured to
be 153 km consisting 9 interconnected SMFs. An optical amplifier with 20
dB gain is used at the input to the 3rd, 6th, and 8th SMF, respectively.
A 32 GBaud PDM-16QAM signal carried by a 1550 nm laser light is launched
with 0 dBm power into the fiber loop and coherently detected at the
receiver end. We use a 80 GSa/s oscilloscope to sample the detected
signal and run signal processing offline. Fig. \ref{fig:exp_cd} shows
the long term CD estimation results when (\ref{eq:cdxx}) and
(\ref{eq:cdxy}) are used. It confirms the stability of (\ref{eq:cdxy})
by comparing Fig. \ref{fig:exp_cd}(a) and (c) where polarization
scrambling is enabled at the fiber input. The maximal SOP speed is
limited to 2000 rad/s by the device in our lab. Note that without
scrambling, the natural condition of the field cable is fairly stable as
shown in Fig. \ref{fig:exp_cd}(b). A 1-hour BER test is performed and
the results are shown in Fig. \ref{fig:exp_ber}. It is again confirmed
that the TED $-\operatorname{Im}\{\operatorname{det}(\bar{P'})\}$ gives
the best performance for all cases.

\section{Conclusion}
We have studied in detail the CD and PMD effect on the second-order
cyclostationarity-based timing recovery commonly used in optical
coherent receivers. We have proposed for $\alpha = 1/T_0$ the estimator
of CD
\begin{align}
\widehat{DL} = \frac{cT_0}{\lambda^2}\cdot\tau_\text{CD}
\end{align}
where $\tau_\text{CD} = \arg\max
\big\{|R_{xx}^\alpha(\tau)|^2 + |R_{yx}^\alpha(\tau)|^2\big\}$,
the estimator of PMD matrix
\begin{align}
\label{eq:ut}
\widehat{U_T} = \hat{C}(\tau_\text{CD}) \approx e^{-j2\pi\alpha\tau_g}U
\,,
\end{align}
where $\tau_g$ is the group delay, the DGD estimator
\begin{align}
\widehat{\tau_\text{DGD}} = 
\big|\arg\{\rho_1\rho_2^*\}\big|/(2\pi\alpha)
\,,
\end{align}
where $\rho_1$ and $\rho_2$ are the two eigenvalues of $\widehat{U_T}$,
and the PSP estimator
$\vec{p}=[\widehat{p_1},\widehat{p_2},\widehat{p_3}]$ with
\begin{align}
\widehat{p_i} = k\cdot\operatorname{Im}\left[
\operatorname{trace}\left(
\sigma_i\widehat{U_T}
\right)\cdot
\operatorname{trace}
(\widehat{U_T})^H
\right]\,,
\end{align}
where $\sigma_i$ is one of the three Pauli matrices. The CD estimator is
insensitive to polarization rotation, PMD and timing error. The PMD
matrix estimator is immune to polarization rotation but affected by
timing error. The DGD and PSP estimators are not affected by
polarization rotation and timing error, but are valid only when the true
DGD is less than $T_0/2$.

Based on the estimators, we have proposed timing error detectors for
the case when the DGD is negligible
\begin{align}
e_T = \operatorname{Im}\operatorname{trace}(\widehat{U_T})
\,,
\end{align}
and for the case when DGD is smaller than $T_0/2$
\begin{align}
e_T = \operatorname{Im}\operatorname{trace}(\widehat{U_T}\hat{U}^H)
\,,
\end{align}
where $\hat{U} = \cos(\pi\alpha\widehat{\tau_\text{DGD}})I 
-j\sin(\pi\alpha\widehat{\tau_\text{DGD}})[\vec{p}\cdot\vec{\sigma}]$
, and for larger DGD to use the adaptive combination of 
$R_{xx}^\alpha$, $R_{xy}^\alpha$, and $R_{yy}^\alpha$,
and finally for all channel conditions
\begin{align}
\label{eq:last}
e_T = \operatorname{det}(\widehat{U_T})
\,.
\end{align}
Both numerical simulations and field tests confirm that (\ref{eq:last})
is a universal timing phase detector regardless of the CD and PMD
condition when the matrix is given by (\ref{eq:ut}).

% Can use something like this to put references on a
% page by themselves when using endfloat and the
% captionsoff option.
\ifCLASSOPTIONcaptionsoff
	\newpage
\fi

\bibliographystyle{IEEEtran}
\bibliography{IEEEabrv,mybib}

\begin{thebibliography}{10}
\providecommand{\url}[1]{#1}
\csname url@samestyle\endcsname
\providecommand{\newblock}{\relax}
\providecommand{\bibinfo}[2]{#2}
\providecommand{\BIBentrySTDinterwordspacing}{\spaceskip=0pt\relax}
\providecommand{\BIBentryALTinterwordstretchfactor}{4}
\providecommand{\BIBentryALTinterwordspacing}{\spaceskip=\fontdimen2\font plus
\BIBentryALTinterwordstretchfactor\fontdimen3\font minus
  \fontdimen4\font\relax}
\providecommand{\BIBforeignlanguage}[2]{{%
\expandafter\ifx\csname l@#1\endcsname\relax
\typeout{** WARNING: IEEEtran.bst: No hyphenation pattern has been}%
\typeout{** loaded for the language `#1'. Using the pattern for}%
\typeout{** the default language instead.}%
\else
\language=\csname l@#1\endcsname
\fi
#2}}
\providecommand{\BIBdecl}{\relax}
\BIBdecl
\renewcommand{\BIBentryALTinterwordstretchfactor}{4}

\bibitem{savory2010digital}
S.~J. Savory, ``Digital coherent optical receivers: Algorithms and
  subsystems,'' \emph{IEEE Journal of Selected Topics in Quantum Electronics},
  vol.~16, no.~5, pp. 1164--1179, 2010.

\bibitem{faruk2017digital}
M.~S. Faruk and S.~J. Savory, ``Digital signal processing for coherent
  transceivers employing multilevel formats,'' \emph{Journal of Lightwave
  Technology}, vol.~35, no.~5, pp. 1125--1141, 2017.

\bibitem{kikuchi2015}
K.~Kikuchi, ``Fundamentals of coherent optical fiber communications,''
  \emph{Journal of Lightwave Technology}, vol.~34, no.~1, pp. 157--179, 2015.

\bibitem{zhou2014eff}
X.~Zhou, ``Efficient clock and carrier recovery algorithms for single-carrier
  coherent optical systems: A systematic review on challenges and recent
  progress,'' \emph{IEEE Signal Processing Magazine}, vol.~31, no.~2, pp.
  35--45, 2014.

\bibitem{godard1978}
D.~Godard, ``Passband timing recovery in an all-digital modem receiver,''
  \emph{IEEE Transactions on Communications}, vol.~26, no.~5, pp. 517--523,
  1978.

\bibitem{gardner1986}
F.~Gardner, ``A {BPSK/QPSK} timing-error detector for sampled receivers,''
  \emph{IEEE Transactions on Communications}, vol.~34, no.~5, pp. 423--429,
  1986.

\bibitem{oerder1988}
M.~Oerder and H.~Meyr, ``Digital filter and square timing recovery,''
  \emph{IEEE Transactions on Communications}, vol.~36, no.~5, pp. 605--612,
  1988.

\bibitem{diniz2018clock}
J.~C.~M. Diniz, F.~Da~Ros, and D.~Zibar, ``Clock recovery challenges in
  dsp-based coherent single-mode and multi-mode optical systems,'' \emph{Future
  Internet}, vol.~10, no.~7, p.~59, 2018.

\bibitem{hauske2010impact}
F.~Hauske, N.~Stojanovic, C.~Xie, and M.~Chen, ``Impact of optical channel
  distortions to digital timing recovery in digital coherent transmission
  systems,'' in \emph{2010 12th International Conference on Transparent Optical
  Networks}.\hskip 1em plus 0.5em minus 0.4em\relax IEEE, 2010, pp. 1--4.

\bibitem{zibar2011exp}
D.~Zibar, J.~C.~R. de~Olivera, V.~B. Ribeiro, A.~Paradisi, J.~C. Diniz, K.~J.
  Larsen, and I.~T. Monroy, ``Experimental investigation and digital
  compensation of {DGD} for 112 {Gb/s PDM-QPSK} clock recovery,'' \emph{Optics
  Express}, vol.~19, no.~26, pp. B429--B439, 2011.

\bibitem{huang2014perf}
L.~Huang, D.~Wang, A.~P.~T. Lau, C.~Lu, and S.~He, ``Performance analysis of
  blind timing phase estimators for digital coherent receivers,'' \emph{Optics
  Express}, vol.~22, no.~6, pp. 6749--6763, 2014.

\bibitem{wang2021modified}
D.~Wang, Z.~Su, H.~Jiang, G.~Liang, Q.~Zhan, and Z.~Li, ``Modified square
  timing error detector with large chromatic dispersion tolerance for optical
  coherent receivers,'' \emph{Optics Express}, vol.~29, no.~13, pp.
  19\,759--19\,766, 2021.

\bibitem{malouin2012natural}
C.~Malouin, P.~Thomas, B.~Zhang, J.~O'Neil, and T.~Schmidt, ``Natural
  expression of the best-match search {Godard} clock-tone algorithm for blind
  chromatic dispersion estimation in digital coherent receivers,'' in
  \emph{Signal Processing in Photonic Communications}.\hskip 1em plus 0.5em
  minus 0.4em\relax Optica Publishing Group, 2012, pp. SpTh2B--4.

\bibitem{sun2011novel}
H.~Sun and K.-T. Wu, ``A novel dispersion and {PMD} tolerant clock phase
  detector for coherent transmission systems,'' in \emph{2011 Optical Fiber
  Communication Conference and Exposition and the National Fiber Optic
  Engineers Conference}.\hskip 1em plus 0.5em minus 0.4em\relax IEEE, 2011, pp.
  1--3.

\bibitem{stojanovic2012}
N.~Stojanovi{\'c}, C.~Xie, Y.~Zhao, B.~Mao, and N.~G. Gonzalez, ``A circuit
  enabling clock extraction in coherent receivers,'' in \emph{2012 38th
  European Conference and Exhibition on Optical Communications}.\hskip 1em plus
  0.5em minus 0.4em\relax IEEE, 2012, pp. 1--3.

\bibitem{rozental2017}
V.~N. Rozental, B.~Corcoran, and A.~J. Lowery, ``Correlation-based polarization
  demultiplexing for clock recovery in coherent optical receivers,'' in
  \emph{Optical Fiber Communication Conference}.\hskip 1em plus 0.5em minus
  0.4em\relax Optica Publishing Group, 2017, pp. W2A--47.

\bibitem{sun2015clock}
H.~Sun, ``Clock and carrier recovery for coherent receivers,'' in \emph{2015
  European Conference on Optical Communication (ECOC)}.\hskip 1em plus 0.5em
  minus 0.4em\relax IEEE, 2015, pp. 1--3.

\bibitem{gardner1986role}
W.~Gardner, ``The role of spectral correlation in design and performance
  analysis of synchronizers,'' \emph{IEEE Transactions on Communications},
  vol.~34, no.~11, pp. 1089--1095, 1986.

\bibitem{ionescu2015cyclo}
M.~Ionescu, M.~Sato, and B.~Thomsen, ``Cyclostationarity-based joint monitoring
  of symbol-rate, frequency offset, cd and osnr for nyquist wdm
  superchannels,'' \emph{Optics Express}, vol.~23, no.~20, pp.
  25\,762--25\,772, 2015.

\bibitem{yan2013digital}
M.~Yan, Z.~Tao, L.~Dou, L.~Li, Y.~Zhao, T.~Hoshida, and J.~C. Rasmussen,
  ``Digital clock recovery algorithm for {Nyquist} signal,'' in \emph{2013
  Optical Fiber Communication Conference and Exposition and the National Fiber
  Optic Engineers Conference (OFC/NFOEC)}.\hskip 1em plus 0.5em minus
  0.4em\relax IEEE, 2013, pp. 1--3.

\bibitem{stojanovic2014}
N.~Stojanovic, B.~Mao, and Y.~Zhao, ``Digital phase detector for {Nyquist} and
  faster than {Nyquist} systems,'' \emph{IEEE Communications Letters}, vol.~18,
  no.~3, pp. 511--514, 2014.

\bibitem{moeneclaey1990}
M.~Moeneclaey and T.~Batsele, ``Carrier-independent {NDA} symbol
  synchronization for {M-PSK}, operating at only one sample per symbol,'' in
  \emph{[Proceedings] GLOBECOM'90: IEEE Global Telecommunications Conference
  and Exhibition}.\hskip 1em plus 0.5em minus 0.4em\relax IEEE, 1990, pp.
  594--598.

\bibitem{stojanovic2011}
N.~Stojanovic, F.~N. Hauske, C.~Xie, and M.~Chen, ``Clock recovery in coherent
  optical receivers,'' in \emph{Photonic Networks, 12. ITG Symposium}.\hskip
  1em plus 0.5em minus 0.4em\relax VDE, 2011, pp. 1--4.

\bibitem{kikuchi2011clock}
K.~Kikuchi, ``Clock recovering characteristics of adaptive
  finite-impulse-response filters in digital coherent optical receivers,''
  \emph{Optics Express}, vol.~19, no.~6, pp. 5611--5619, 2011.

\end{thebibliography}

% \section*{Biographies} 
% \begin{footnotesize} 
% Dawei Wang (wangdw9@mail.sysu.edu.cn) is an associate
% professor with the School of Electronics and
% Information Technology at Sun Yat-sen University,
% Guangzhou, China. He received the B.S. and Ph.D.
% degree in optical engineering from Zhejiang
% University, Hangzhou, China in 2008 and 2013,
% respectively. He has six years working experience in
% Huawei Technologies, Shenzhen, China. His research
% interests include digital signal processing and
% optical performance monitoring. \newline

% Qi Sui (sui-qi@hotmail.com) received the B.Eng.
% degree from Shanghai Jiaotong University, Shanghai,
% China, in 2007, and Ph.D. degree from the Hong Kong
% Polytechnic University, Hung Hom, Hong Kong, in 2015.
% He is with the Institute of Photonics Technology,
% Jinan University, China. His research interest
% includes the optical communications and optical
% performance monitoring.
% \newline

% Zhaohui Li (lzhh88@mail.sysu.edu.cn) is a professor
% in the School of Electronics and Information
% Technology at Sun Yat-sen University, China. He
% obtained his B.S. from the Department of Physics and
% his MSc from the Institute of Modern Optics of Nankai
% University, China, in 1999 and 2002, respectively,
% and received his Ph.D. from Nanyang Technological
% University in 2007. His research interests include
% optical communication systems, optical signal
% processing technology and ultrafine measurement
% systems.
% \vfill
% \end{footnotesize}

\end{document}